\documentclass[12pt,preprint,eqsecnum,aps]{revtex4}
\usepackage{graphicx}
\usepackage{amssymb}
\begin{document}

\title{Quantum particle statistics on the holographic screen leads to Modified Newtonian Dynamics  (MOND)}

\author{E. Pazy and N. Argaman}

\affiliation{Department of Physics, NRCN, P.O.B. 9001, Beer-Sheva 84190, Israel}

\begin{abstract}
Employing a thermodynamic interpretation of gravity based on the holographic principle and assuming underlying particle statistics, fermionic or bosonic, for the excitations of the holographic screen leads to Modified Newtonian Dynamics (MOND). A connection between the acceleration scale $a_0$ appearing in MOND and the Fermi energy of the holographic fermionic degrees of freedom is obtained. In this formulation the physics of MOND results from the quantum-classical crossover in the fermionic specific heat. However, due to the dimensionality of the screen, the formalism is general and applies to two dimensional bosonic excitations as well. It is shown that replacing the assumption of the equipartition of energy on the holographic screen by a standard quantum-statistical-mechanics description wherein some of the degrees of freedom are frozen out at low temperatures is the physical basis for the MOND interpolating function ${\tilde \mu}$. The interpolating function ${\tilde \mu}$ is calculated within the statistical mechanical formalism and compared to the leading phenomenological interpolating functions, most commonly used. Based on the statistical mechanical view of MOND, its  cosmological implications are re-interpreted: the connection between $a_0$ and the Hubble constant is described as a quantum uncertainty relation; and the relationship between $a_0$ and the cosmological constant is better understood physically.
\end{abstract}

\maketitle

\section {Introduction }
\label {sec:Introduction}

The connection between gravity and thermodynamics was first noted in the
pioneering works of Bekenstein \cite{Bekenstein73} and Hawking \cite{Hawking75} on black hole thermodynamics. Later on the idea was further expanded by Unruh \cite{Unruh76} who identified the connection between acceleration and temperature, demonstrating that an accelerating observer will observe a black-body radiation whose temperature would be proportional to his acceleration. Employing these ideas and turning the line of argument around, Jacobson \cite{Jacobson95}  derived the Einstein field equations from the laws of thermodynamics, based on the assumption that the proportionality between area and entropy, derived by Bekenstein for black holes, is universal. Similar results were also obtained by Padmanabhan in a series of works reviewed in \cite{Padmanabhan10}. Verlinde \cite{Verlinde10} introduced the idea that Newton's law of gravitation can be understood as an entropic force, basing this result on the holographic approach and the thermodynamical formulation of gravity; similar ideas were also presented by Padmanabhan  \cite{Padmanabhan04}.

On the other hand, seemingly unrelated to the thermodynamic interpretation of gravity, gravitational theory is faced with observational challenges. Observational discrepancies between the observed mass in a galaxy and its galactic rotation curves and large velocities in galaxy clusters are already long standing problems. Attempts to solve this observational puzzle have resulted in the introduction of "Dark Matter" as well as alternative gravity theories such as Modified Newtonian Dynamics (MOND) \cite{Milgrom83}. In the late 1990's, a second cloud appeared in the horizon, when observations of distant red shift relations indicated that the expansion of the universe is accelerating \cite{dark_energy}, implying a positive cosmological constant $\Lambda$. The idea of a cosmological constant was first introduced by Einstein himself, as it appears naturally in his field equations. However problems arise when considering its observed physical value and the attempt to connect it with  the quantum mechanical vacuum energy.
MOND, introduced ad hoc to solve discrepancies on the galactic scale, has also had success in explaining observations regarding superclusters \cite{Milgrom97}. However it seemed to have no cosmological predictions. It is thus surprising to find out that the acceleration scale $a_0$ introduced into MOND to phenomenologically explain the observed galaxy rotation curves, is related to the value of the Hubble constant, $H_0$, through the relationship  $(a_0/2 \pi) \approx c H_0$ \cite{Milgrom83a} and to the cosmological constant as well $(a_0)  \approx c{(\Lambda/3)}^{1/2}/2 \pi$ \cite{Milgrom99}.

These seemingly unconnected views of gravity, i.e. MOND and the thermodynamic approach are nevertheless interlinked, and obtaining an underlying microscopic theory for them will help explain the cosmological aspects of MOND. From the thermodynamic representation of gravity it seems natural to relate the constant $a_0$, having dimensions of acceleration, to a temperature via the Unruh relationship, resulting in a temperature scale. Based on Verlinde's idea of gravity being an entropic force, several attempts have been recently made to obtain MOND by considering some of the degrees of freedom on the holographic screen to be frozen out. In \cite{Li10} MOND was obtained by considering a one dimensional Debye model for the excitations on the holographic screen thus restricting the excited degrees of freedom at low temperatures, whereas \cite{Kiselev11} considered collective excitations on the holographic screen thus obtaining MOND. MOND was also obtained by considering a minimal temperature on the holographic screen \cite{Klinkhamer2011} and relating it to $a_0$. In \cite{Neto10} a non-homogenous cooling of the holographic screen was considered resulting from a phase transition occurring at a critical temperature; under this assumption a modified Friedmann equation compatible with MOND theory was also obtained. This work was followed and extended by \cite{Hendi10} in which entropic corrections to the theory where considered. The  work in \cite{Modesto10} should also be noted for obtaining MOND through entropic volume corrections to Newton's law. The present work simply assumes that degrees of freedom on the holographic screen should be treated through the quantum-statistical-mechanics formalism; following this assumption we not only obtain MOND but we are able to calculate its interpolating function ${\tilde \mu}$ and compare it to leading phenomenological expressions which are based on astronomical data.

In Sec. \ref{Sec:Verlinde} Verlinde's thermodynamic formulation of gravitation is briefly described. In Sec. \ref{Sec:Verlinde_mod} a modification to Verlinde's theory is introduced by replacing the equipartition rule for excitations on the holographic screen, by the quantum statistical mechanical expression for the energy of a fermionic or bosonic two dimensional gas. Via this replacement MOND is obtained and the connection of the MOND interpolating function ${\tilde \mu}$ to the two dimensional specific heat is established. The obtained interpolating function is then compared to the MOND phenomenological interpolating functions. Cosmological implications of the statistical mechanical interpretation of $a_0$ are described in Sec. \ref{sec:cosmological}. A short summary is then given in Sec. \ref{sec:summary}.

\section {A formulation of the thermodynamic theory of gravity}
\label {Sec:Verlinde}
The connection between gravity and thermodynamics has been greatly developed by
Padmanabhan  \cite{Padmanabhan04} and Verlinde \cite{Verlinde10}. In this section we choose to describe this connection through a formulation introduced by Verlinde.
We start by briefly introducing this formulation following section 3 of his paper \cite{Verlinde10}, which is based on four well known essential equations from which one obtains Newtonian gravity theory.

Consider a point mass, $M$, surrounded by a spherical holographic screen of radius $R$. Thermodynamics on the holographic screen is connected to gravitation  by applying two equations. The first is the Unruh relation between the temperature, $T$ and the acceleration $a$ of an observer at the screen,
\begin{equation}
k_BT={1 \over 2 \pi }{\hbar a \over c},
\label{eq:Unruh}
\end{equation}
where $c$ is the speed of light and $k_B$ is Boltzmann's constant. For simplicity of notation we shall use energy units such that $k_B=1$. The second relationship is the relation obtained by Bekenstein for the number of bits or degrees of freedom, $N$, on the Horizion of a Black hole, which Verlinde extends to the holographic screen,
\begin{equation}
N={A c^3 \over G \hbar},
\label{eq:Bekenstein}
\end{equation}
where $A$ is the area of the holographic screen and $G$ is Newton's gravitational constant. The two remaining equations needed to complete the model are Einstein's mass, energy, $E$, relation
\begin{equation}
E=Mc^2
\label{eq:Einstein}
\end{equation}
and the thermodynamic equipartition rule
\begin{equation}
E={1 \over 2} N T.
\label{eq:equipartition}
\end{equation}
It should be noted that in this approach it is the gravitational energy of Eq. (\ref{eq:Einstein}) which is related to thermal excitations on the holographic screen.

Combining these four equations (\ref{eq:Unruh}-\ref{eq:equipartition}), and expressing the holographic screen area by its radius, $A=4 \pi R^2$, one directly obtains Newton's law of gravitation,
\begin{equation}
a={G M \over R^2}.
\label{eq:Newton}
\end{equation}

\section {Quantum statistical extension}
\label {Sec:Verlinde_mod}
In this section we extend Verlinde's model, introduced in the previous section, by considering the quantum particle statistics of the bits on the holographic screen. The equipartition rule described in Eq. (\ref{eq:equipartition}) is considered as the Dulong Petite law, valid for the high temperature limit, which needs to be modified at lower temperatures due to the underlying particle statistics. It will be shown that considering the quantum statistical nature of the excitations on the holographic screen leads to MOND (for a recent review of MOND theory see \cite{MOND}).

\subsection {The physical interpretation of $a_{0}$ in terms of Fermionic excitations}
\label {subsection:fermionic}

We start by considering fermionic excitations, on the holographic screen.
In defining the particle statistics of fermions one needs to introduce an energy scale,
the Fermi energy $E_{F}$, which distinguishes between excited thermal states and states which are "frozen out". In the case of fermionic excitations of the holographic screen, this energy scale will be related to $a_0$. In Verlinde's thermodynamic gravitational formulation, gravitational effects are related only to
the thermal excitations of the holographic screen, as can be deduced from the equipartition rule (\ref{eq:equipartition}). Thus in considering gravitational effects the
systems ground state energy should be ignored and one should consider only thermal excitations.

We begin by calculating the energy $E_{tot}$, of an excited two dimensional fermionic system due to the heating of the system to a temperature $T$, and use the expression obtained to replace the equipartition relation (\ref{eq:equipartition}). The energy of the two dimensional Fermi gas is obtained by calculating the following integral \cite{LL},
\begin{equation}
E_{tot}= {g A m \over 2 \pi {\hbar}^2}\int_0^{\infty} {\epsilon d\epsilon \over \exp{[(\epsilon -\mu)/T}]+1},
\label{eq:energy_low_temp}
\end{equation}
where,  $g=2s+1$, $s$ is the spin of the particle, $m$, is its mass, and $\mu$ is the chemical potential. The energy was denoted as $E_{tot}$ to distinguish it from the energy appearing in Eqs. (\ref{eq:Einstein}, \ref{eq:equipartition}), which is the gravitational energy related only to thermal excitations. Calculating the integral to second order in the temperature one obtains
\begin{equation}
E_{tot}= E_0+ {g A m \pi \over 12 {\hbar}^2} T^2.
\label{eq:energy}
\end{equation}
The number of  particles in the system, is given by
\begin{equation}
N_{par}={g A m \over 2 \pi {\hbar}^2}\int_0^{\infty} { d\epsilon \over \exp{[(\epsilon -\mu)/T}]+1},
\label{eq:number}
\end{equation}
In our derivation we consider the particle number, $N_{par}$, to be the free variable whereas $\mu$ is determined through Eq. (\ref{eq:number}).
In the zero temperature limit $T=0$, one obtains
\begin{equation}
N_{par}^0={g A m \over 2 \pi {\hbar}^2}E_F,
\label{eq:particlenum}
\end{equation}
where $E_F$ is the system's Fermi energy. Finite temperature corrections to the particle number are exponentially small in $T/E_F$.
We can now express the system's thermal energy, $E= E_{tot}-E_0$, in terms of the temperature, the particle number and the Fermi energy,
\begin{equation}
E= {T^2 N_{par}^0 {\pi}^2 \over 6 E_F}.
\label{eq:energy_th}
\end{equation}
The above expression for the thermal energy replaces the equipartition relation (\ref{eq:equipartition}).

Employing Eq. (\ref{eq:energy_th}), we follow Verlinde's steps using the three remaining equations (\ref{eq:Unruh}-\ref{eq:Einstein}) and some algebraic manipulations
to obtain MOND.  We start by obtaining an expression for the temperature squared,
\begin{equation}
T^2={6 M c^2E_F \over N_{par}^0{\pi}^2}.
\label{eq:temperature}
\end{equation}
Relating the temperature to the acceleration through the Unruh formula
(\ref{eq:Unruh}), we obtain
\begin{equation}
a^2={24 c^2  \over {\hbar}^2} {M c^2 E_F \over N_{par}^0}.
\label{eq:acceleration}
\end{equation}
In two dimensions $N_{par}=N/2$ where $N$ is the number of degrees of freedom and is equal to the number of Planck cells on the holographic screen. Thus
 \begin{equation}
N_{par}^0={Ac^3 \over 2 G \hbar}.
\label{eq:number_0}
\end{equation}
The area of the screen is given by $A=4 \pi R^2$. Eq. (\ref{eq:acceleration}) is very similar to the MOND equation in the deep MOND limit
\begin{equation}
a ({a \over a_0})=G {M \over R^2}.
\label{eq:MOND}
\end{equation}
The MOND equation (\ref{eq:MOND}) is obtained, employing equation Eq.(\ref{eq:number_0}) and identifying $a_0$ as
\begin{equation}
a_0={12 c \over \hbar \pi }E_F.
\label{eq:a0}
\end{equation}

The Fermi energy defines an energy scale, relating it to an acceleration scale the same way temperature is transformed, Eq. (\ref{eq:Unruh}) we define
$E_F=(\hbar / 2 \pi ) ({\tilde E_F} / c)$
and obtain
\begin{equation}
a_0={\tilde E_F \over b},
\label{eq:a0_norm}
\end{equation}
where $b\equiv \pi^2 / 6$.
It should be noted that the Newtonian limit for Eq. (\ref{eq:MOND}) is obtained at the high temperature (acceleration) limit since the Maxwell-Boltzmann statistics is the high temperature limit of the Fermi, Bose statistics. When $T>>E_F$
the fermionic particle distribution in Eq. (\ref{eq:energy_low_temp})
goes to the Maxwell-Boltzmann limit and the limit is independent of $E_F$ i.e. $a_0$.

\subsection {The interpolating function ${\tilde \mu}$}
\label{subsection:interpolating function}

\begin{figure}
\centering
\includegraphics{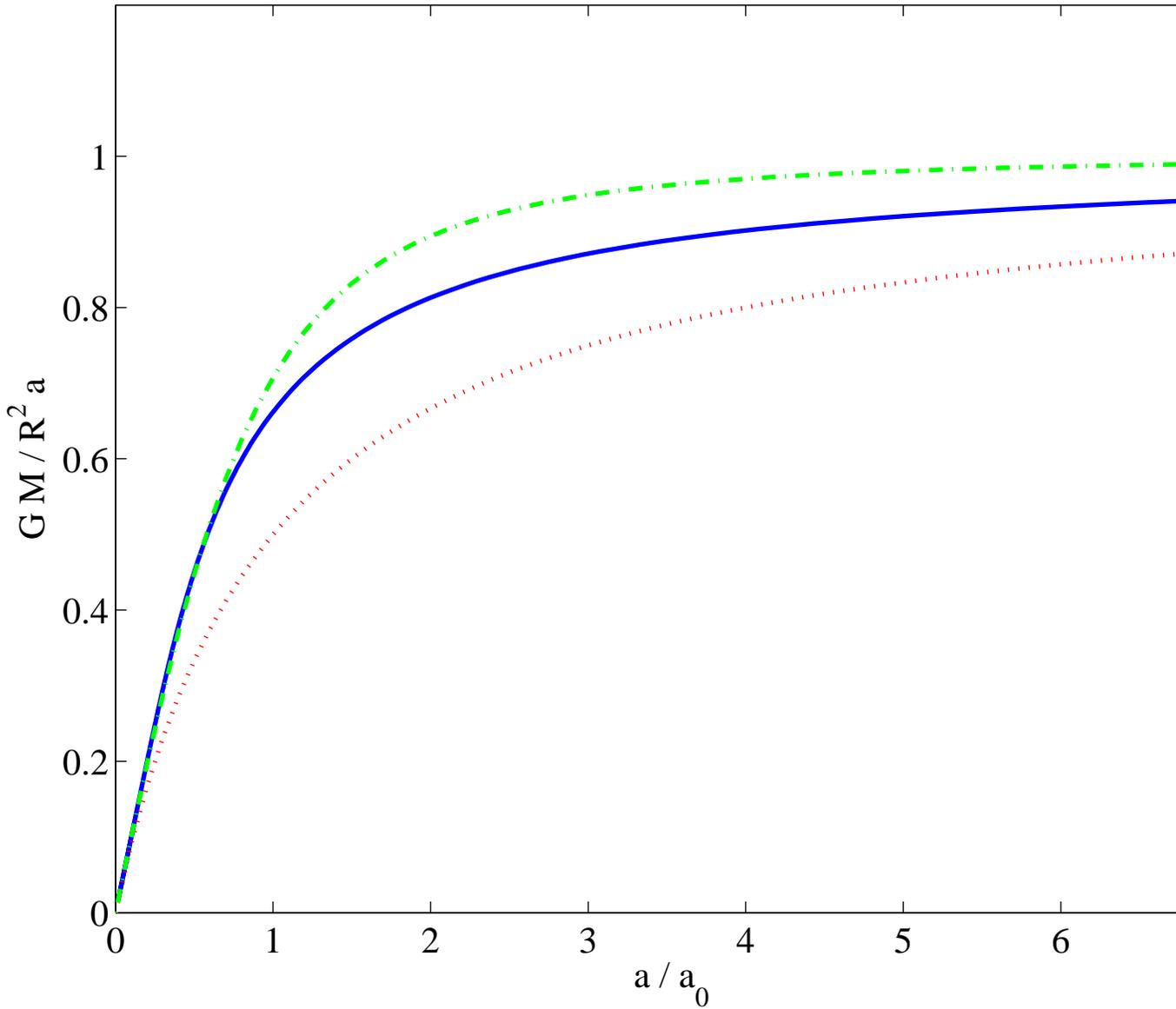}
\caption{Comparison of ${\tilde \mu}$ as obtained by the statistical mechanical
considerations (full line) to two of the leading MOND interpolating functions: the "standard" (dot-dashed line)
 ${\tilde \mu}(x>>1)={x / \sqrt{1+x^2}}$ and the "simple" (dotted line) ${\tilde \mu}(x>>1)={x /(1+x)}$ interpolating functions.}
\label{fig1}
\end{figure}

Regarding the physical interpretation of Eq. (\ref{eq:a0_norm}) which relates $a_0$ to
${\tilde E_F}$, it should be noted that since the Fermi energy is related to the density of the particles, $a_0$ can also be viewed as a constant inter-particle distance on the holographic screen. We obtained the above correspondence for $a_0$ by introducing fermionic degrees of freedom on the holographic screen and considering the deep MOND regime, i.e., very low accelerations, $a<<a_0$. The high temperature regime was shown to correspond to the Newtonian limit. In the intermediate regime MOND is characterized by an interpolating function ${\tilde \mu}$ which defines the MOND formula $ \vec{a} {\tilde\mu } ({|\vec{a}|/a_0})=- \vec{\nabla} \Phi$. The asymptotic behavior of the function
 ${\tilde \mu}(x); x=a/a_0$, in the low acceleration regime $x \rightarrow 0$
 is ${\tilde \mu}(x)=x$ corresponding to the deep MOND limit, and in the high acceleration limit ${\tilde \mu}(x)=1$, defining the Newtonian limit described above. Whereas in MOND the interpolating function is obtained phenomenologically from astronomical data, we can use our statistical mechanical interpretation in terms of the underlying fermionic degrees of freedom to obtain ${\tilde \mu}(x)$ in the intermediate regime. We start by expressing
 ${\tilde \mu}(x)$ in MOND as the following ratio
 \begin{equation}
\left ({GM \over R^2}\right )/a={\tilde \mu}(a/a_0).
 \label{eq:mu_ratio}
 \end{equation}
 Employing eqs. (\ref{eq:Unruh}),(\ref{eq:Bekenstein}) and (\ref{eq:Einstein}) we can write the above ratio as
 \begin{equation}
 {E \over N_{par} T}={\tilde \mu}(a/a_0).
 \label{eq:mu_ratio_Eth}
 \end{equation}
 $E$ can be calculated from Eq. (\ref{eq:energy_low_temp}) under the constraint of a fixed particle number given by Eq. (\ref{eq:number}). Subtracting from the result the ground state energy one obtains the thermal energy, $E$. We have performed this calculation numerically  and the result is expressed in Fig. (\ref{fig1}) by the continuous line. As expected the function crosses over from a linear dependence for small $a$ (low temperatures) to a $1-c/a$ dependence for large $a$ (high temperature). The result is compared with two leading phenomenological expressions for the MOND ${\tilde \mu}$ function,
 $ {\tilde \mu}=x/(1+x)$, known as the "simple" $\mu$-function which is expressed in the figure as a dot-dashed line and ${\tilde \mu}=x/\sqrt{(1+x^2)}$, which is also commonly used \cite{MOND}, known as the "standard" interpolating function, designated in the figure by the dotted line. Both interpolating functions belong to the $n-$family of interpolating functions ${\tilde \mu}=x/(1+x^n)^{1 /n}$, where the $n=1$ describes the "simple" interpolating function and the $n=2$ the "standard" interpolating function. It should however be noted that these MOND interpolating functions are put in by hand, whereas the function in Eq. (\ref{eq:mu_ratio_Eth}) is a result of physical considerations.

 The data on galaxy rotation curves is becoming more and more restrictive regarding which functions can be considered as reasonable interpolating functions. Nowadays the data seem to favor the "simple", $n=1$ interpolating function or some interpolation between $n=1$ to the "standard" interpolating function $n=2$ \cite{MOND}. The thermodynamic interpolating function we have calculated seems to do exactly that.

It should be noted that the same calculation for the thermal energy $E$ performed numerically to obtain ${\tilde \mu}$ in Eq. (\ref{eq:mu_ratio_Eth}) can be performed analytically, and the result can be expressed in terms of the dilog function $Li_2(y)$
\begin{equation}
\label{eq:eth_analytical}
E=-{N_{par}^0\over E_F}[T^2 Li_2(-e^{\mu/T})+{E_F^2 \over 2}]
\end{equation}
where $N_{par}^0$ and $E_F$ are given in Eq.(\ref{eq:particlenum}) and $\mu$ is defined through Eq. (\ref{eq:number}). Thus an analytical expression can be given for the ${\tilde \mu}(x)$ MOND interpolation function
\begin{equation}
\label{eq:mu_analytical}
{\tilde \mu}(a/a_0)=-{b\over a a_0}[\left({a \over  b}\right)^2 Li_2(-e^{{\bar \mu}/a})+ a_0^2],
\end{equation}
where $\bar \mu=(\hbar / 2 \pi ) ({\tilde \mu} / c)$ is the chemical potential related  to an acceleration scale the same way temperature is transformed, Eq. (\ref{eq:Unruh}).
The MOND interpolating function, ${\tilde \mu}$, in the thermodynamic interpretation is simply the thermal energy divided by the total number of excitations times the temperature, thus it can be viewed as the relative number of the thermal excitations.

In the low temperature limit ${\tilde \mu}$ can be connected to the specific heat for the two dimensional fermionic gas. To demonstrate this connection we compare the thermal energy to $E=Mc^2$ but in this case we do not estimate the integral as was done in Eq. (\ref{eq:energy}), instead to obtain the thermal energy we use the specific heat integrating it up to a given temperature
\begin{equation}
\int_{0}^{T} dT' C_V ({T' \over E_F})=Mc^2
\label{eq:mu_general}
\end{equation}
where the partial derivative was replaced by the specific heat $C_V=({\partial E / \partial T})_{V}$. The lowest order term corresponding to the zero temperature case, the ground state energy being irrelevant, we are left with the leading order expression. Since for low temperatures the specific heat is linear in the temperature we can obtain $E$ in terms of the low temperature specific heat, thus
\begin{equation}
\left({T \over 2}\right) C_V ({T \over E_F})\cong Mc^2.
\label{eq:mu_der}
\end{equation}
From Eq. (\ref{eq:mu_ratio_Eth}) we identify ${\tilde \mu}$ in the MOND equation with the specific heat, divided by the temperature times $N_{par}^0$ the number of degrees of freedom, obtaining the following relation
\begin{equation}
{1 \over N_{par}^0}C_V ({T \over E_F }) \cong {\tilde \mu}({a \over a_0}).
\label{eq:mu_specific_heat}
\end{equation}

The physical interpretation of Eq. (\ref{eq:mu_specific_heat}) is straight-forward: applying a force to a body, in trying to accelerate the body we are also attempting to heat degrees of freedom on the holographic screen, our ability to do so is given by the specific heat of the screen. However, the physical basis for MOND is revealed in Eq. (\ref{eq:mu_ratio_Eth}),
which shows that the interpolating MOND function $\tilde \mu$ is essentially the relative number of thermal excitations since it is given by the ratio of the thermal excitation energy divided by the high temperature thermal excitation energy, where in this limit each degree of freedom gets an energy of $T/2$.

In the high temperature limit the physics does not depend on the quantum nature of excitations. Rather each excited degree of freedom receives an energy of ${T / 2} $ as defined by the equipartition rule. In this limit we can simply follow Verlinde's formulation and obtain Newtonian dynamics. The Newtonian limit in the formulation of MOND obtained by taking the limit $a_0 \rightarrow 0$, has a simple physical meaning, in the formulation of MOND via the specific heat, Eq. (\ref{eq:mu_specific_heat}) the Newtonian limit results directly from the Dulong Petite law. Even though the high temperature limit is governed by the Dulong Petite law obtaining the first asymptotic correction to ${\tilde \mu}$ in the high temperature limit $T > T_0$ is not straightforward. ${\tilde \mu}$ is proportional to the ratio between the thermal energy and the temperature
(\ref{eq:mu_ratio_Eth}) in the Dulong Petite law the energy is linear in temperature
however there is also a temperature independent part to the energy as can be deduced
from Eq. (\ref{eq:eth_analytical}); this term gives a correction to ${\tilde \mu}$ which is inverse in the temperature. Numerically one obtains ${\tilde \mu}(x>>1) \approx 1- (0.41/x)$, in the particle statistics formulation whereas employing the "simple" interpolating function the asymptotic correction is ${\tilde \mu}_{simple}(x>>1) \approx 1- (1/x)$ and via the "standard" interpolating function one obtains ${\tilde \mu}_{standard}(x>>1) \approx 1- (1/2x^2)$. In general, for the  $n-$family of interpolating functions the asymptotic correction is given by ${\tilde \mu}_n(x>>1) \approx 1- (1/nx^n)$.

\subsection{Bosonic extension}
\label {subsec:Bosonic}

The above relationship (\ref{eq:mu_specific_heat}) between the two dimensional specific heat and ${\tilde \mu}$ in the MOND equation was obtained for fermions, however in two dimensions the specific heat for an ideal gas of Fermi particles is identical to the specific heat of an ideal Bose gas for all $T$ and $N$. Thus in general the acceleration $a_0$ is related to the temperature $T_0$ which divides the classical from the quantum regime. The physical meaning of  $({a_0 / a})$ is obtained by the connection to thermodynamics in which $({T / T_0})^{1/2}$ is the ratio of the mean interparticle separation to the thermal wavelength \cite{May64}.

Since the typical temperature scale, $T_0$, separating the classical from the quantum regime is identified with the MOND acceleration scale $a_0$, our result applies both to Fermi as well as Bose excitations of the holographic screen. The result for bosons is expressed in terms of $T_0$, instead of in terms of $E_F$. It should be realized that the temperature scale $T_0$ does not correspond to a critical temperature associated with a phase transition; quite the opposite is true. The reason the fermionic and bosonic two-dimensional specific heat can be identical, is  the fact that there is no Bose condensation in two dimensions.

Eq. (\ref{eq:mu_specific_heat}) is valid for the low temperature limit, and was obtained for fermions; to verify it for bosons we first consider the case of a two dimensional bosonic gas composing the holographic screen, and comparing it to our previous results we obtain their equivalence to the fermionic results. We start with the general expression for the two dimensional specific heat, both for fermions and for bosons, \cite{Anghel02}
\begin{equation}
C_V(y_0)=-{N_{par}^2 \over T \sigma}{1+y_0 \over y_0}-2T\sigma Li_2(-y_0)
\label{eq:Cv_lt}
\end{equation}
where $N_{par}$ is the total number of particles, $\sigma$ is a constant defined as $\sigma= {g A m/4\pi \hbar^2}$ and $y_0$ is defined through the relationship $N_{par} =T \sigma \log(1+y_0)$ which holds for fermions as well as for bosons.
Since, $y_0 \rightarrow \infty$ when $T \rightarrow 0$, we obtain from Eq. (\ref{eq:mu_der}) in the low temperature limit where $Li_2(-y_0)\approx -[{\pi^2 / 6}+ \log^2(y_0)/2]$ the expression
\begin{equation}
T^2={Mc^2 \over \sigma b}.
\label{eq:particl_stat}
\end{equation}
To compare this with our previous results we insert the fermionic expression $\sigma=(N_{par}^0/2 E_F)$ and obtain Eq. (\ref{eq:temperature}), through which the fermionic result in Eq. (\ref{eq:a0}) is also obtained. From Eq. (\ref{eq:mu_specific_heat}), taking the low temperature limit $C_V/N= (b T \sigma /N)$, we obtain via Eqs. (\ref{eq:Unruh}, \ref{eq:a0_norm}) the required MOND low acceleration limit,
\begin{equation}
{\tilde \mu}(x<<1) = x,
\label{eq:mu}
\end{equation}
where $x=(a/a_0)$. It should be noted that the leading order corrections to (\ref{eq:mu}) are exponentially small in $x$.

\section{Cosmological implications}
\label{sec:cosmological}

Having obtained MOND through a quantum statistical mechanical view we proceed to review MOND's cosmological implications through similar considerations. It should be noted that
the key equations defining the connections between the MOND acceleration scale $a_0$ and cosmological scales were all previously obtained. The purpose of this section is mainly
to reinterpret previous results in terms of a quantum statistical mechanical view.

The relationship between the MOND acceleration scale $a_0$ and the Hubble constant
\begin{equation}
{a_0 \over 2 \pi} \approx c H_0,
\label{eq:uncertainity}
\end{equation}
was obtained in observations. It turns out employing the quantum statistical mechanical description the above relation has a simple quantum mechanical interpretation as a cosmological energy time uncertainty relation, $\Delta E \Delta t \approx \hbar$. $ a_0$  relates through the Unruh formula, Eq. (\ref{eq:Unruh}), to the energy dividing quantum and classical regimes, thus relating to an energy uncertainty $\Delta E \approx (\hbar a_0 /2 \pi c )$. The inverse of the Hubble constant relates to a time uncertainty $\Delta t \approx 1/ H_0$, and combing both we obtain a cosmological quantum uncertainty relation Eq. (\ref{eq:uncertainity}).

The second cosmological relationship related to MOND is the connection between $a_0$
and the square root of the cosmological constant \cite{Milgrom99,Ho2011}
\begin{equation}
a_0 \approx {a_{\Lambda}\over 2 \pi} ,
\label{eq:cosmolgical_constant}
\end{equation}
where $a_{\Lambda}=\sqrt{\Lambda/3}$.
Astronomical observations indicate we live in an accelerating universe \cite{dark_energy}, i.e., one defined by a positive cosmological constant $\Lambda>0$. In a cosmological constant dominated universe a connection between the cosmological constant and $a_0$ had been obtained in \cite{Milgrom99} and recently reviewed in \cite{Ho2011}. We briefly review its derivation and use our statistical mechanical interpretation of MOND to explain the connection. The net temperature measured by a non-inertial observer with acceleration, $a$, in a de Sitter universe is given by \cite{Milgrom99},
\begin{equation}
\tilde{T}= (\sqrt{a^2+a_{\Lambda}^2}-a_{\Lambda})/2 \pi,
\label{eq:background}
\end{equation}
which is simply an acceleration analog of the background reference temperature arising due to the universe's acceleration. In \cite{Ho2011} it was shown by considering the limit $a << a_{\Lambda}$ that one obtains Eq. (\ref{eq:cosmolgical_constant}). Similar considerations were also presented in \cite{Klinkhamer2011}. The relationship (\ref{eq:cosmolgical_constant}) has the same physical interpretation in a quantum statistical mechanical description of excitations on the holographic screen. However in the quantum statistical description there is a natural connection between
$a_0$ as the Fermi energy or its bosonic analog to the background reference temperature or energy with respect to which the excitations defined via a non-inertial Unruh temperature are measured.

Through the expression for the background reference temperature (\ref{eq:background}) Milgrom obtained an expression for the MOND interpolating function, ${\tilde \mu}$ \cite{Milgrom99},
\begin{equation}
{\tilde \mu}=[1+(2x)^{-2}]^{1/2}-(2x)^{-1}.
\label{eq:mu_background}
\end{equation}
It is interesting to note that the asymptotic expansion for large $x$ for the above interpolating function(\ref{eq:mu_background}) is ${\tilde \mu}(x>>1) \approx 1- 1/2x$
which is very close to the asymptotic expansion via the quantum statistical approach
${\tilde \mu}(x>>1) \approx 1- (0.41/x)$, whereas the latter was obtained for the
dynamics due to a given mass $M$. However, the asymptotic expansion for small $x$
of the interpolating function(\ref{eq:mu_background}) has corrections of the order
of $O(x^3)$ to the leading $x$ term whereas for the quantum statistical approach
we obtained corrections which are exponentially small in $1/x$, i.e., $O(\exp{[-1/x]})$.
Unfortunately these differences are extremely small on all relevant scales and thus are
almost impossible to discriminate with current astronomical observations.

\section{summary}
\label {sec:summary}
In summary, a quantum-mechanical microscopic description has been found to lead to MOND. Through this approach the physics of MOND has been shown to arise from the possibility of creating excitations on the holographic screen; more specifically for low temperatures it was directly related to the specific heat of fermionic or bosonic excitations of the holographic screen. The MOND acceleration term, $a_0$, was first shown to correspond to the Fermi energy of excitations on the holographic screen; later it was shown to apply also to bosonic excitations, thus corresponding more generally to a temperature scale $T_0$, separating the classical from the quantum regime. A general expression for the  MOND interpolating function was obtained and its physical meaning was shown to be related to the relative number of
thermal excitations on the holographic screen, which in turn can be related to the temperature integral of the specific heat or directly to the specific heat for low temperatures. Moreover the interpolation function was calculated numerically and compared with leading phenomenological interpolating functions. The calculated quantum statistics based interpolation function seems to fit well with the best estimated phenomenological MOND interpolation functions. It is thus important to stress that the quantum mechanical microscopic basis approach is not only a physical basis for MOND; it is a physical theory with observable predictions. Even though the interpolating function arising from the theory seems to agree with the leading phenomenological functions, there still are some differences in high-order corrections. Whereas corrections to the linear leading order term in the phenomenological interpolating functions, in the deep MOND regime, i.e. small $a/a_0$, are polynomial in $a/a_0$ in the quantum mechanical microscopic description these corrections are exponentially small in $a_0/a$, i.e., $O(\exp{[-1/x]})$.

On the cosmological scale the relationship between $a_0$ and the Hubble constant was shown to be related to an energy time uncertainty and $a_0$ was shown to correspond to the background reference Unruh temperature arising from the universe's acceleration.

E. P. would like to thank I. Gurwich for discussions.

\newpage

\begin{thebibliography}{99}

\bibitem{Bekenstein73}
J. D. Bekenstein Phys. Rev. D {\bf 7}, 2333 (1973).

\bibitem{Hawking75}
S. W. Hawking, Commun. Math. Phys. {\bf 43}, 199 (1975).

\bibitem{Unruh76}
W. Unruh, Phys. Rev. D {\bf 14}, 870 (1976).

\bibitem{Jacobson95}
T Jacobson, Phys. Rev. Lett {\bf 75}, 1260 (1995).

\bibitem{Padmanabhan10}
T. Padmanabhan, Rept. Prog. Phys. {\bf 73}, 046901 (2010).

\bibitem{Verlinde10}
E. Verlinde, J. High Energy Phys. 29 (2011) 04; arXiv:1001.0875.

\bibitem{Padmanabhan04}
T. Padmanabhan, Classical Quant. Grav {\bf 21}, 4485 (2004);
T. Padmanabhan, arXiv:0912.3165.

\bibitem{Milgrom83}
M. Milgrom, Astrophysics. J. {\bf 270}, 371 (1983).

\bibitem{dark_energy}
A. G. Riess, {\it et al.}, Astronomical. J. {\bf 116}, 1009 (1998);
S. Perlmutter, {\it et al.}, Astrophysics. J. {\bf 517}, 565 (1999).

\bibitem{Milgrom97}
M. Milgrom, Astrophysics. J. {\bf 480}, 7 (1997).

\bibitem{Milgrom83a}
M. Milgrom, Astrophysics. J. {\bf 270}, 365 (1983).

\bibitem{Milgrom99}
M. Milgrom, Phys. Lett. A {\bf 253}, 273 (1999).

\bibitem{Li10}
X. Li and Z. Chang, Commun. Theor. Phys, {\bf 55} 733
(2011) ; arXiv:1005.1169.

\bibitem{Kiselev11}
 V. V. Kiselev and S. A. Timofeev, Mod. Phys. Lett. A {\bf 26}, 109 (2011).

\bibitem{Klinkhamer2011}
F. R. Klinkhamer and  M. Kopp, Mod. Phys. Lett. A {\bf 26}, 2783 (2011); arXiv:1104.2022.

\bibitem{Neto10}
J. A. Neto, arXiv:1009.4944.

\bibitem{Hendi10}
S. H. Hendi and A. Shekhi, Rev. D {\bf 83}, 084012 (2011); arXiv:1012.0381.

\bibitem{Modesto10}
L. Modesto and A. Randono, arXiv:1003.1998.

\bibitem{MOND}
B. Famacy and S. McGaugh, arXiv:1112.3960.

\bibitem{LL}
L. D. Landau and E. M. Lifshitz, {\it Statistical Physics}. Oxford Pergamon (1989).

\bibitem{May64}
R. M. May, Phys. Rev {\bf 135}, 1515 (1964).

\bibitem{Anghel02}
D. V. Anghel, J. Phys A: Math Gen {\bf 35}, 7255 (2002).

\bibitem{Zhao}
H. S. Zhao and B. Famey, Astrophys. J. {\bf 638}, 9 (2006).

\bibitem{Ho2011}
 C. M. Ho, D. Minic and Y. J. Ng, arXiv:1105.2916.

\end{thebibliography}
\end{document}